# Room-temperature self-cavity lasing from organic color centers

Minna Zhang[1], Hao Wu[1]*, Xuri Yao[1], Jiyang Ma[1], Mark Oxborrow[2], and Qing Zhao[1]*

1.Center for Quantum Technology Research and Key Laboratory of Advanced Optoelectronic Quantum Architecture and Measurements (MOE), School of Physics, Beijing Institute of Technology Beijing 100081, China
2.Department of Materials Imperial College London South Kensington, London SW7 2AZ, UK

E-mail: hao.wu@bit.edu.cn; qzhaoyuping@bit.edu.cn

**Abstract**

Color centers, which are point defects in crystals, play a crucial role in altering the optical properties of their host materials, enabling widespread applications in the field of quantum information processing. While the majority of the state-of-the-art color centers are inorganic, they come with limitations such as the challenging material preparations and insufficient amount of available centers. In contrast, organic color centers have recently gained attention due to their ease of preparations and tailorable functionalities. Here, pentacene-doped *p*-terphenyl (Pc:Ptp), an organic color-center system normally used for microwave quantum electronics, is demonstrated for the first time its ability of self-cavity laser emission at room temperature. The laser emission is characterized by strong polarization and high anisotropy, attributed to the unique packing of the color-center molecules within the crystal. The optical coherence is found to be a figure of merit to distinguish the processes of the amplified spontaneous emission (ASE) and lasing in Pc:Ptp. This work highlights the potential of Pc:Ptp as a compact and efficient platform for light-matter interactions , offering significant promise for enhancing the performance of solid-state quantum devices based on this organic color-center system.

**Introduction**

Color centers, as point defects in crystals, enrich many properties of the crystal systems, especially the quantum-state-dependent absorption and/or emission characteristics,[1–5] enabling widespread applications in quantum information processing[1–6] and quantum sensing[1–6]. At present, the inorganic color centers, e.g., nitrogen-vacancy (NV) centers in diamond,[2,3,5] silicon vacancies in silicon carbide,[4] fluorine centers in alkali halide crystals[1,7] are mainly used due to their excellent optical and spin properties at room temperature. For instance, NV color centers in diamond have unique electron spin states that can be manipulated using magnetic fields, microwaves, or light.[1–3,8] Their spin states can be initialized, optically read, and coherently controlled. Thus, they can be applied to metrology[1,3], magnetometry,[1,9–11] pressure sensing[12] and thermometry.[13] Long spin coherence time and well-defined optical transitions at room temperature make it suitable for applications in quantum computing[3,14–20] and quantum communication.[3,21–29]

Among above quantum applications, the efficiency of the optical readout of color centers is crucial for the device implementation. To this end, increasing the concentration of the color centers is one of the most common solutions.[6,9,30,31] However, the high-concentration crystals with satisfying material qualities are challenging to be prepared.[3,6] Besides, it has been found that too high concentration (e.g., >0.1 ppm of nitrogen concentration in diamond) can decrease the coherence time of the color centers,[6] thus, degrade the device performance. Alternatively, the exploitation of the lasing process in the color centers has become a promising strategy for enhancing the optical readout efficiency owing to the boosted emission intensity as well as the narrowed emission linewidth.[8,32,33] The recent attempts have successfully demonstrated the stimulated emission[8,32,34] of the inorganic color centers while the lasing is rare due to the high threshold, insufficient gain medium and the requirement of the sophisticated optical feedback architecture.[8,33,34]

In addition to inorganic color centers, recent years have witnessed rapid growth of the interest in organic color centers, which possess the intriguing advantage of the tailorable functionalities for quantum applications.[35–40] As one of the representative organic color centers, pentacene-doped *p*-terphenyl (Pc:Ptp) can circumvent the shortcomings of the inorganic systems. Pc:Ptp is a low-cost, easy-to-manufacture organic single crystal that can be produced in a bulk size with high quality.[41] Due to the substituent doping, pentacene molecules, as functional species, can be well protected by the *p*-terphenyl matrix. Pc:Ptp has the excellent photophysical properties not found in pure pentacene that well suitable for the state-of-the-art room-temperature quantum devices.[42–44] Examples include masers,[45] quantum memory,[42] spin refrigerators[43], logic qubits[44] and quantum magnetometers.[46] These applications mainly utilize the photoexcited triplet states of pentacene in *p*-terphenyl. Very recently, the amplified spontaneous emission (ASE) of Pc:Ptp in the visible light band[47–49] exploiting the pentacene's singlet transition has been observed, suggesting the feasibility of its stimulated emission at ambient conditions. However, the potential of the lasing in Pc:Ptp remains unexplored. The realization of the Pc:Ptp laser is a significant step towards improving the emerging quantum sensors[46,50] relying on the optical readout of the pentacene's singlet states.

In this paper, we demonstrate for the first time the self-cavity laser emission from Pc:Ptp at room temperature. The inherent structure of the millimeter-sized Pc:Ptp crystal forms a Fabry–Pérot (FP) cavity naturally facilitating the laser oscillation. The laser output is highly polarized with a polarization ratio of 0.747, and demonstrates anisotropic emission related to the long-range molecular order and the transition dipole moments (TDMs). Moreover, the coherence features (rarely investigated in organic solid-state lasers[51]) of the laser emission from Pc:Ptp were comprehensively characterized, showing the power for distinguishing the ASE and lasing processes of which the criteria are normally ambiguous[52–59]. Our work demonstrates the application potential of the Pc:Ptp color centers in the field of solid-state lasers. We envision such lasers can be employed to develop nascent organic quantum sensors by combining with the simultaneously formed hyperpolarized triplet spins of pentacene.

**Results and Discussion**
**Self-cavity lasing**

Pink Pc:Ptp single crystals with a doping concentration of 1000 ppm were used. The color centers are the pentacene molecules replacing the *p*-terphenyl molecules in two inequivalent sites of the monoclinic lattice as shown in **Fig. 1a**. More specifically, the pentacene molecules are arranged almost standing on the crystal ab plane, as illustrated in **Fig. 1b**. The angle between the long axis of the pentacene molecule (i.e., the x-axis) and the ab plane is 72.78°, and the short in-plane molecular axis (i.e., the y-axis) of pentacene is almost parallel to the ab plane.[41] The "head-to-head" molecular stacking on the crystal ab plane results in weaker intermolecular interactions than the π-π interactions between the molecular xy planes, leading the ab plane to be the cleavage plane of the crystal. By gently cutting the sample, square-millimeter-size Pc:Ptp thin flakes can be obtained with the naturally formed cleavage surfaces.

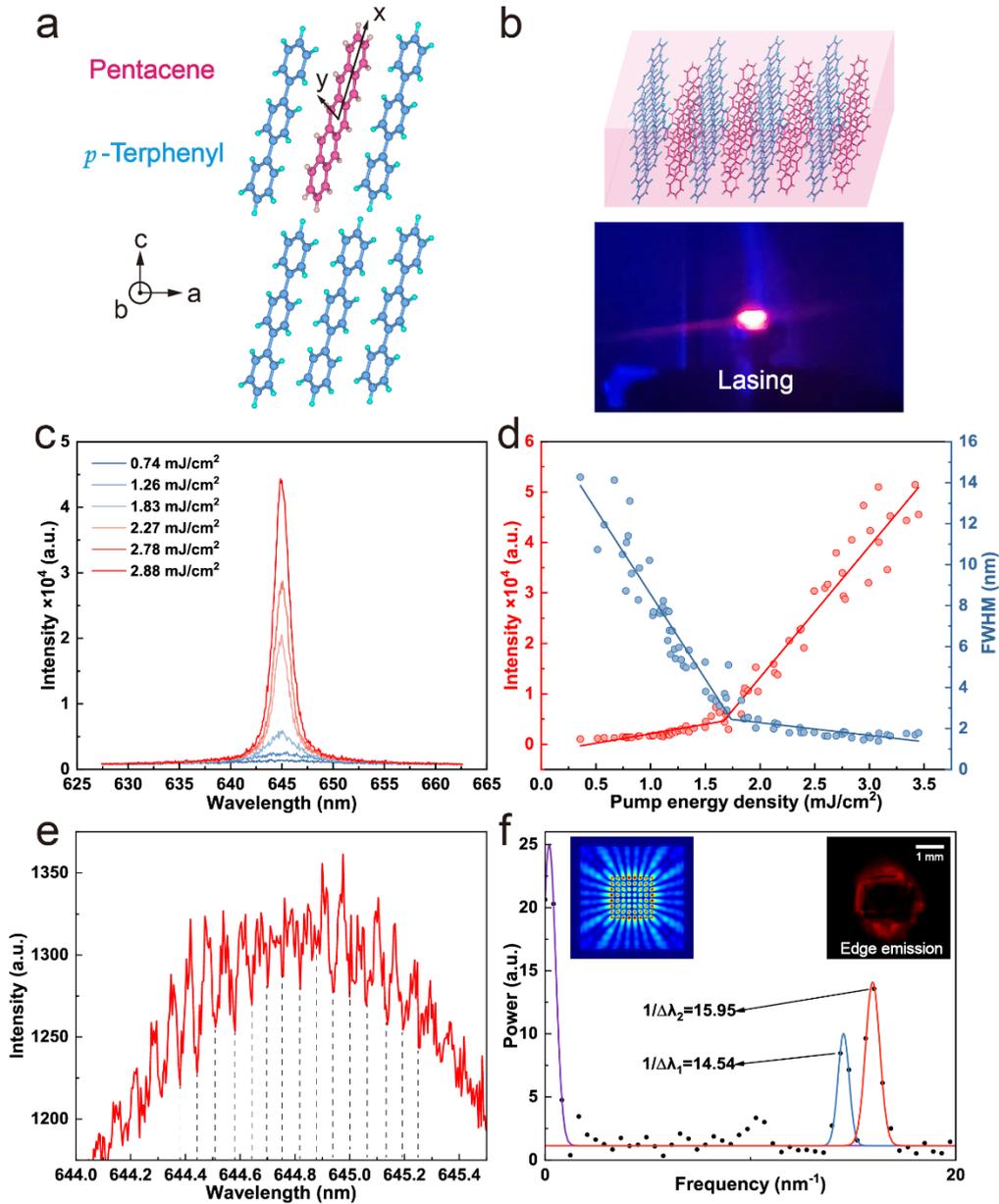

**Fig. 1 Crystal structure and optical emission of Pc:Ptp. a**) Crystal structure of Pc:Ptp. The crystal axes (a, b, and c) as well as the molecular axes (x and y) are labelled. **b**) Top: molecular packing of the pentacene and *p*-terphenyl molecules within the crystal where the large surface plane is the crystal ab plane. Bottom: the photograph of the lasing of Pc:Ptp under the optical pumping. The pump light is filtered. **c**) Effect of the pump energy density on the emission spectrum of Pc:Ptp. **d**) Dependence of the intensity (red) and full width at half maximum (FWHM) (blue) of the emission peak at 645 nm on the pump energy density. Both datasets are fit with a piecewise function for obtaining the pump threshold. **e**) Zoom-in on the emission peak at 645 nm measured above the pump threshold. **f**) The fast Fourier transform (FFT) of the spectrum in **e**). The two FFT peaks are fit by Lorentzian functions. Top left: the simulated Fabry–Pérot mode within the Pc:Ptp crystal. Top right: the photograph of the laser emission from the edge of the crystal with the pump light filtered.

By pumping Pc:Ptp with a 590-nm laser perpendicular to the cleavage surface, strong red emission with a certain directivity can be observed as shown in **Fig. 1b**. **Fig. 1c** reveals the dependence of the emission spectrum on the pump energy density that the stronger pumping gives rise to the higher emission intensity as well as the spectral narrowing at 645 nm. By plotting the intensity and the full width at half maximum (FWHM) of the emission peak as a function of the pump energy density, the process of the stimulated emission is confirmed by the presence of the pump threshold at 1.7 mJ cm$^{-2}$, as shown in **Fig. 1d**. Further, **Fig. 1e** shows that the narrowed emission peak obtained with the pump energy density above the threshold in fact contains multiple spikes with uniform mode intervals. Two reciprocal mode intervals, 14.54 and 15.95 nm$^{-1}$, were obtained by the fast Fourier transform (FFT) (see **Fig. 1f**) of the spikes, which corresponded to the mode intervals about 0.06-0.07 nm. Such spikes have also been observed in other organic solid-state lasers,[51,57,60] indicating the coherent feedback is likely to exist in the crystal, resulting in the laser oscillation. Different from the ASE process, the lasing process requires a cavity to introduce the optical feedback.[61] For the organic crystals with well-defined crystal surfaces and high refractive indices, the crystals by themselves can behave like optical cavities, giving rise to the self-cavity lasers. The most common cavity formed by the crystal itself is the FP cavity, which uses two parallel crystal facets as mirrors.[51,60] In addition, the crystals with specific shapes[56,58] can act as the whispering-gallery mode (WGM) cavities, in which light propagates along the cavity-surrounding interface via total internal reflection, leading to the much higher quality ($Q$) factors and smaller mode volumes compared to those of FP cavities.

To clarify the origin of the self-cavity lasing in Pc:Ptp, we have combined the experimental results with theoretical investigations. For the FP and WGM cavities discussed above, there are[51,62]

$$m\lambda = 2nd = \frac{mc}{\nu} \tag{1}$$

$$\Delta\nu = \frac{c}{2nd} \tag{2}$$

$$\Delta\lambda = \frac{\lambda^2}{2nd} \tag{3}$$

where m (an integer) is the order of the mode, $\lambda$ is the cavity wavelength, n is the refractive index of the material, d is the length of the cavity, $\nu$ is the cavity frequency, $\Delta\nu$ is the frequency interval of the mode, $\Delta\lambda$ is the wavelength interval of the mode. For the Pc:Ptp crystal used, there were three potential optical cavities that could form naturally: (i) the WGM cavity formed by the internal reflections along the four edge facets (**Fig. S3**, left, **Supplementary Information**); (ii) the FP cavity formed by the two opposite edge facets; (iii) the FP cavity formed by the largest facets, i.e., the cleavage surfaces (**Fig. S3**, right, Supplementary Information). Given the size of the Pc:Ptp crystal was 2 mm×2 mm×0.25 mm, the three cases correspond to $d_i$ = 2.83 mm, $d_{ii}$ = 2 mm, $d_{iii}$ = 0.25 mm. Let λ = 645 nm, n = 1.606, three Δλ values can be calculated respectively according to **Equation (3)**, which are 0.046 nm, 0.065 nm and 0.518 nm. The second value agrees well with the mode interval obtained by the FFT of the emission spikes in **Fig. 1e,f**, indicating the rationality of the hypothesis (ii). The reasons for the two FFT peaks in **Fig. 1f** may arise from the non-uniform cavity length defined by the imperfect edge facets and the birefringence of *p*-terphenyl.[63] The simulation of the field distribution of the corresponding FP mode is shown in **Fig. 1f**. Indeed, even with the straightforward visual inspection, we found the strong red emission always appeared near the edges of the sample as shown in **Fig. 1f** while almost no emission was from the largest facets. We further characterized the *Q* factor of the FP cavity formed by the sample itself based on the definition $Q = \lambda/\text{FWHM}$[58]. The *Q* factor was calculated to be 464.

**Lasing polarization and directionality**

Prior to investigating the polarization and directionality of the Pc:Ptp laser, we determined the crystal a and b-axes within the cleavage surface (i.e. the crystal ab plane) for optimizing the laser output using two approaches. Given the a and b-axes are orthogonal and there is no birefringence along the crystal a-axis,[64] by monitoring the birefringence phenomenon of the crystal with an optical microscope (see **Methods**), both axes can be determined respectively as shown in **Fig. 2a**. Alternatively, by pumping the sample with a laser whose polarization is tunable and parallel to the cleavage surface, the crystal b-axis can be found which is aligned with the laser polarization giving the maximum photoluminescence.[49] The above approaches mutually confirmed the crystal axes on the cleavage ab plane and in the following experiments, the polarization of the pump laser was always kept parallel to the crystal b-axis for maximizing the transition probability and thus optimizing the laser emission intensity.

The polarization comparison of the optical emission from the Pc:Ptp crystal below and above the lasing threshold is shown in **Fig. 2b**. The polarized (laser) emission is only achieved when the pump energy density is above the threshold. The Pc:Ptp laser emission has a high polarization contrast of about 0.75 due to the optical anisotropy of the crystal,[52–55,59,65–74] which is related to the ordered arrangement of the molecules.[53,65,75,76] It is defined by the TDM of the $S_1 \rightarrow S_0$ electron transition, and the resulting exciton cause the crystal's emission to be polarized.[52–55,59,65–74,77–80] Meanwhile, the molecular arrangement in organic crystals follows the π-stack[81] or herringbone structure.[81] The main structure of Pc:Ptp crystal belongs to the herringbone structure of Fig.7-c in ref.81. Exciton coupling is generally much larger within layers than between layers,[82] so π-stacks and herringbone arrangements without translational invariance result in partial depolarization.[82] Conversely, the long-range ordered herringbone structure with translation invariance is the reason for the high polarization of Pc:Ptp crystals.

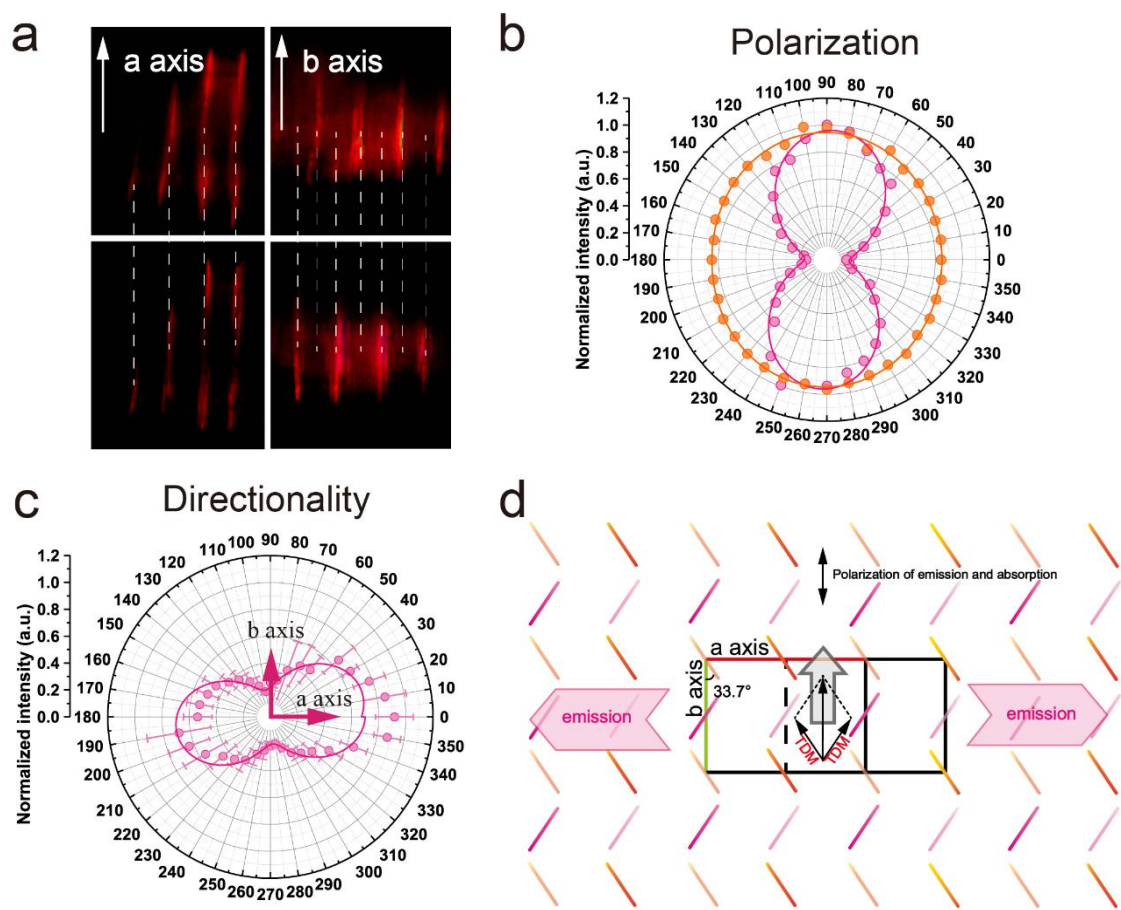

**Fig. 2 Polarization and directionality of the Pc:Ptp laser. a**) The birefringence of the Pc:Ptp crystal observed by an optical microscope. The red lines are the scales of a ruler placed underneath the crystal. The dash lines are used to indicate the alignment between the observed scales before (top) and after (bottom) rotating the polarizer of the microscope by 90°. No birefringence shows along the a-axis (left) to which the b-axis is orthogonal and reveals birefringence (right). **b**) The polarization diagrams of the optical emission from Pc:Ptp below (orange) and above (pink) the lasing threshold. **c**) The directionality of the laser emission from Pc:Ptp. The pink dots are averaged results of five independent measurements and fit with a sinusoidal function. Error bars represent the standard deviations. **d**) Schematic diagram of the projection of the in-plane short axes (i.e. molecular y-axes) of pentacene (pink) and *p*-terphenyl (orange) onto the crystal ab plane of Pc:Ptp. The transition dipole moments (TDMs) of the pentacene molecules doped in two inequivalent sites aligning with the pentacene molecular y-axes are labelled. The direction of the vector sum of the TDMs is parallel to the crystal b-axis.

With the coupling fiber fixed on a homebuilt circular rail shown in **Fig. S1** (see Supplementary Information), the directionality of the Pc:Ptp laser emission was investigated. As indicated in **Fig. 2c**, the Pc:Ptp laser reveals the remarkable directionality and it is noteworthy that the dominated emission direction is parallel to the crystal a-axis and the emission intensity is the weakest along the crystal b-axis.

The relationship between the emission directionality and TDM is shown in Fig. 2d. The gray arrow is the maximum direction of the TDM projection vector. However, the maximum emission falls in the direction perpendicular to the TDM (i.e., the a-axis). This phenomenon can be explained by the fact that under optical pumping, molecules are more easily excited and emitted in the direction of TDM (i.e., polarization along the b axis),[53,83–85] causing light (i.e., electromagnetic waves) to propagate along the other orthogonal direction (i.e., a-axis).

**Lasing coherence**

The coherence of the Pc:Ptp laser was investigated based on the Michelson interference experiment of which the setup is schematically demonstrated in **Fig. S2** (Supplementary Information). The optical fringes, arising from the interference between the split Pc:Ptp optical output beams, under different optical path differences (OPDs) and pump energy densities were collected and quantified based on the fringe visibilities:

$$V = \frac{I_{max} - I_{min}}{I_{max} + I_{min}}, \tag{4}$$

where $I_{max}$ and $I_{min}$ are the maximum and minimum light intensities extracted from the interference fringes (see **Fig. S5** in Supplementary Information for the intensity profiles processed from the fringe patterns). As shown in **Fig. 3**, the fringe visibility increases with the pump energy density while decreases with the OPD. When the pump energy density (i.e. ~0.7 mJ cm$^{-2}$) is far below the lasing threshold even at zero OPD, the interference fringe is almost invisible. By increasing the pump energy density to be above 1.5 mJ cm$^{-2}$, the fringes start to become visible, and the associated visibilities follow a linear increasing trend with the pump energy density.

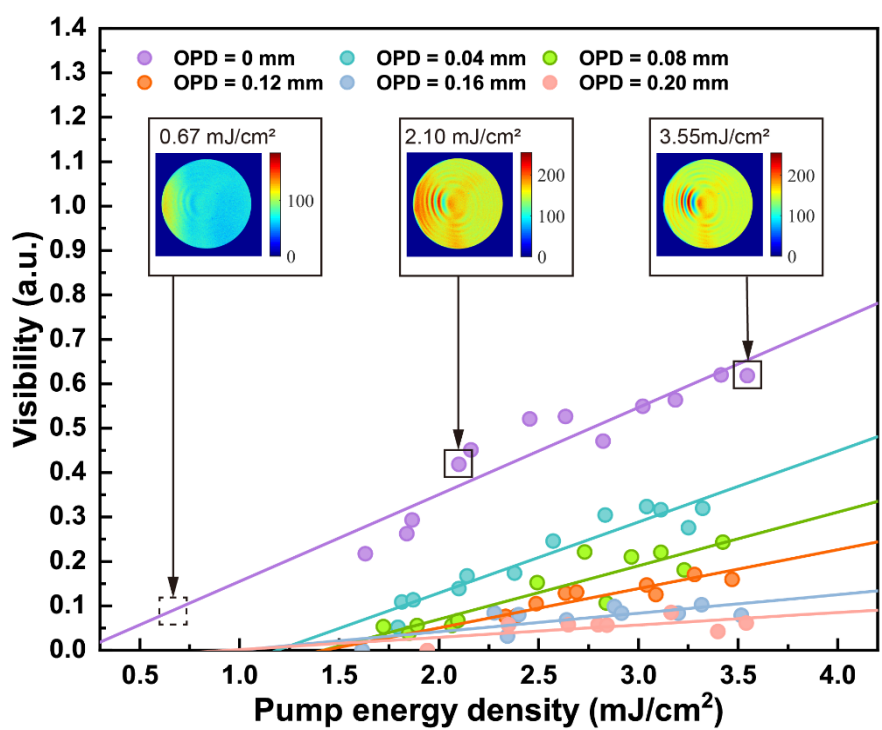

**Fig. 3 The dependence of the fringe visibility on the pump energy density under different OPDs.** The correlation between the fringe visibility and the pump energy density is indicated by the linear fitting. Inset: the fringe patterns collected under different pump energy densities at zero OPD. The dashed box denotes the low visibility of the fringe obtained with the pump energy density far below the lasing threshold.

As the OPD can be converted to the time delay $\Delta t$ between the arms of the Michelson interferometer according to OPD = $c\Delta t$, the dependence of the fringe visibility on the time delay can be obtained and employed to evaluate the coherence time of the Pc:Ptp laser. **Fig. 4a** shows the visibility curves measured under different pump conditions. In order to mitigate the effect of the pump energy fluctuations on quantifying the coherence time, we plot the averaged visibilities of the fringes obtained within specific intervals of the pump energy density as a function of the time delay. The coherence times $\tau_{cFWHM}$ at different pump conditions can be obtained from the FWHMs of the visibility curves in **Fig. 4a**. It can be clearly seen that the stronger pumping gives rise to the broadening of the visibility curves (i.e. larger FWHM), indicating the longer coherence time. The quantitative results are shown in **Fig. 4b** where the coherence time is increased from 0.19±0.02 ps to the maximum 0.37±0.03 ps by enhancing the pump energy density. Thus, a maximum coherence length of 0.11mm is obtained based on $l_{cFWHM} = c\tau_{cFWHM}$. To the best of our knowledge, there are no previous studies on the organic *self-cavity* lasers implementing the coherence measurements which results in the lack of the literature values for fair comparison. On the other hand, the peak visibility rises from 0.25 to 0.6 and an abrupt enhancement is found upon the pump energy density is above 2 mJ cm$^{-2}$.

Furthermore, we have investigated the correlations between the coherence and emission spectral features of Pc:Ptp. As illustrated in **Fig. 4b**, both the results obtained from the coherence and emission spectral measurements reveal the phase-transition-like points indicating the abrupt changes of the associated parameters, i.e., the coherence time, interference visibility, emission intensity and emission linewidth. Notably, for each type of the measurements, there are always two distinct threshold values of the pump energy density, while the thresholds indicated by the fringe peak visibility and coherence time are almost identical to those found in the plots of the emission intensity and linewidth, respectively. It is understandable since the fringe visibility is correlated with the light intensity according to **Equation (4)**. Meanwhile, the coherence time and spectral linewidth obey the following relationship: [86]

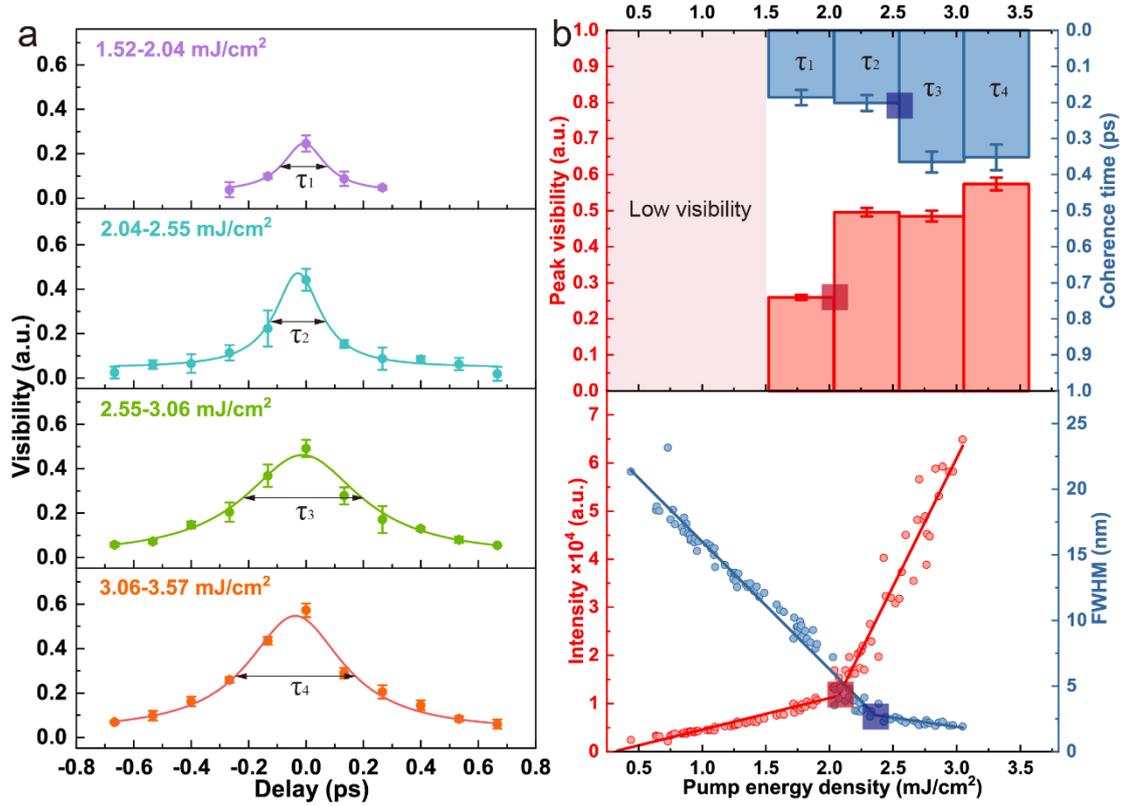

**Fig. 4 Coherence measurements. a**) The dependence of the fringe visibility on the time delay between the arms of the interferometer under different pump-energy-density intervals. The visibility curves are obtained by the Lorentzian fittings of the data points where the FWHM values represent the coherence times ($\tau_1, \tau_2, \tau_3$ and $\tau_4$). **b**) Distinguishment between the Pc:Ptp ASE and lasing processes with the coherence measurements (top) and emission spectral measurements (bottom). The solid red and blue boxes mark the abrupt changes of the associated parameters.

$$\tau_{cFWHM} = \frac{2\ln 2}{\pi c} \frac{\lambda_0^2}{\lambda_{FWHM}} \tag{5}$$

where $\lambda_0$ is the emission wavelength (i.e. ~645 nm), $\lambda_{FWHM}$ is the FWHM of the emission spectral. By inserting the measured coherence times 0.19 ps and 0.37 ps in **Equation (5)**, the theoretical values of the corresponded emission linewidth 3.22 nm and 1.65 nm can be obtained, respectively. In comparison with the measured emission linewidths shown in **Fig. 4b**, the calculated linewidth (3.22 nm) is less than the measured ones (5-10 nm) within the energy density interval (1.5-2.0 mJ cm$^{-2}$). It might be due to the broad background contributed by the spontaneous emission in the spectral measurements with the relative weak pumping, while in the coherence measurements, despite the presence of the incoherent spontaneous emission, it would not contribute to the fringe patterns but lower the fringe visibility as observed. The partial and relatively short coherence of the emission may arise from the onset of the ASE process[87] that is consolidated by the threshold occurring in the emission intensity/peak fringe visibility around 2.0 mJ cm$^{-2}$. In contrast, under the stronger pumping condition (>2.5 mJ cm$^{-2}$), the linewidth (1.65 nm) calculated from the coherence time is much closer to the measured spectral linewidth (2-2.5 nm) which indicates the majority of the emission is coherent. Combing with the observation of the sudden (two-fold) enhancement of the coherence time near 2.5 mJ cm$^{-2}$, the above results imply the emergence of a new phase (i.e. lasing) that dominates the emission and shows a much better coherence compared with the ASE occurring with a lower pump energy density. Thus, the two distinct thresholds in either the coherence or emission spectral measurements can be attributed to the ASE and lasing thresholds, respectively.

**Conclusion**

In summary, Pc:Ptp, as an organic color system, has been successfully demonstrated its capability of self-cavity lasing at room temperature. This development introduces a novel organic solid-state laser gain medium, distinct from those based solely on pure organic systems. The laser exhibits a polarization contrast of 0.75, indicating that the emission from the Pc:Ptp is nearly perfectly linearly polarized. Additionally, the laser emission demonstrates significant directionality (orthogonal to the transition dipole moment), closely related to the molecular arrangement of the color centers. Furthermore, the correlation between the optical emission spectra and coherence of Pc:Ptp reveals a universal methodology for distinguishing ASE from laser emission, a topic that has not been extensively discussed or identified in other organic self-cavity (self-waveguide) lasers. To enhance the utility of emerging organic color center systems like Pc:Ptp in quantum sensing applications (e.g., magnetometry[46]), the laser emission characteristics can be leveraged to improve the optical readout efficiency of the color-center-based quantum sensors[34] while a lower laser threshold and prolonged [i.e. (quasi-)continuous-wave][88] laser emission are highly desired. This can be accomplished by: (i) enhancing the fluorescence quantum yield through optimized doping concentration[36]; (ii) crystal geometry engineering[58,89,90] to improve the $Q$ factor of the optical resonator formed by the gain medium itself; and (iii) addressing the triplet bottleneck problem by regulating the dynamics of the pentacene triplet spins.

# Methods

## Sample Preparation

A Pc:Ptp single crystal with a doping concentration of 1000 ppm was grown using the Bridgman method, as detailed in reference [45]. The crystal was cut to obtain a cleavage facet, which was subsequently polished using abrasive papers, 0.1-µm cerium oxide powder and 0.05-µm aluminum oxide powder. The surface parallel to the finished facet was polished by repeating the aforementioned procedures. The samples used have a width of 1-2 mm and a thickness of 0.25 mm.

## Spectral Measurements

The laser emission of Pc:Ptp was measured using a custom-built setup, as illustrated in **Fig. S1** (Supplementary Information). The sample was positioned at the center of an aluminum alloy circular guide rail. The optical fiber and filter were placed in a tube (not pictured), which was secured to movable slider that moved frictionlessly along the guide rail. The distance between the center of the slider and the center of the sample was 13 cm. For the measurements, an optical parametric oscillator (OPO) (BBOPO-Vis, Deyang Tech. Inc., pulse duration 7 ns) pumped by an Nd:YAG Q-switched laser (Nimma-900, Beamtech, repetition rate 10 Hz) with horizontally polarized output at 590 nm was employed. The OPO output beam was focused on the sample surface using a reverse-placed beam expander (2x) and a convex lens with a focal length of 20 cm. The beam diameter was 5 mm, completely covering the sample surface. A 1/10 beam splitter was employed to divert the pump light for energy measurements using an energy meter (BGS6321, Beijing Institute of Optoelectronic Technology). A long-pass filter with a cut-on wavelength of 600 nm was used to eliminate the pump light. The signals were collected by an optical fiber connected to a high-resolution spectrometer (SpectraPro HRS-750, Teledyne Princeton Instruments). For the spectrum shown in **Fig. 1e**, 100 frames were averaged.

## Axis Identification by Polarizing Microscopy

The crystal was placed on a micrometer, and an optical microscope was used to observe the micrometer underneath the crystal (**Fig. S4** in Supplementary Information). The polarizer was rotated to check for birefringence, and the crystal was subsequently rotated for identifying the a and b axes.

**Coherence Measurements**

A Michelson interferometer setup for measuring coherence length and coherence time is illustrated in **Fig. S2** (Supplementary Information). The stimulated emission from the sample was received by a 7-m long square core fiber (IR400×400-FC/S-1803001-L7) connected to an optical fiber coupler (Beijing Beiguang Century Instrument Co., LTD.), and then transmitted to a 50/50 beam splitter. One of the mirrors was moved to adjust the time delay. The combined light was then directed into a camera (ORCA-Quest qCMOS, C15550-22UP) through a lens (MCX10814-A, N-BK7, ø = 50.8 mm, f = 125.0 mm, AR: 400-700 nm) and an aperture (LBTEK, OPB-12.5B) for capturing the fringe interference pattern of Pc:Ptp emission obtained under different OPDs and pump energy densities.

**Image Processing**

The pseudo-color map of the interference pattern was processed using MATLAB. The light intensity along the same transverse line for each image was plotted. $I_{max}$ and $I_{min}$ were obtained from the fit of each plot to avoid noise effects.

**Data availability**

All data are also available from the corresponding author upon request.

**Acknowledgements**

H.W. acknowledges support from the National Natural Science Foundation of China (12204040), the Beijing Institute of Technology Research Fund Program for Young Scholars (XSQD-6120230016), and the China Postdoctoral Science Foundation (YJ20210035, 2021M700439, and 2023T160049).

**Author contributions**

M.Z., H.W. conceived the experiments and took the lead in writing the manuscript. M.Z. carried out the main experiments. X.Y. helped with the coherence imaging experiment. H.W. and M.O. provided samples. J.M. built the setup for the spectral mesurements. H.W. and Q.Z. supervised the work. All authors provided critical feedback and helped shape the research, analysis and manuscript.

**Competing interests**

The authors declare no competing interests.

**Additional information**

**Supplementary information** The online version contains supplementary material is available.

Supplementary Information

**Room-temperature self-cavity lasing from organic color centers**

Minna Zhang[1], Hao Wu[1]*, Xuri Yao[1], Jiyang Ma[1], Mark Oxborrow[2], and Qing Zhao[1]*

1.Center for Quantum Technology Research and Key Laboratory of Advanced Optoelectronic Quantum Architecture and Measurements (MOE), School of Physics, Beijing Institute of Technology Beijing 100081, China
2.Department of Materials Imperial College London South Kensington, London SW7 2AZ, UK

E-mail: hao.wu@bit.edu.cn; qzhaoyuping@bit.edu.cn

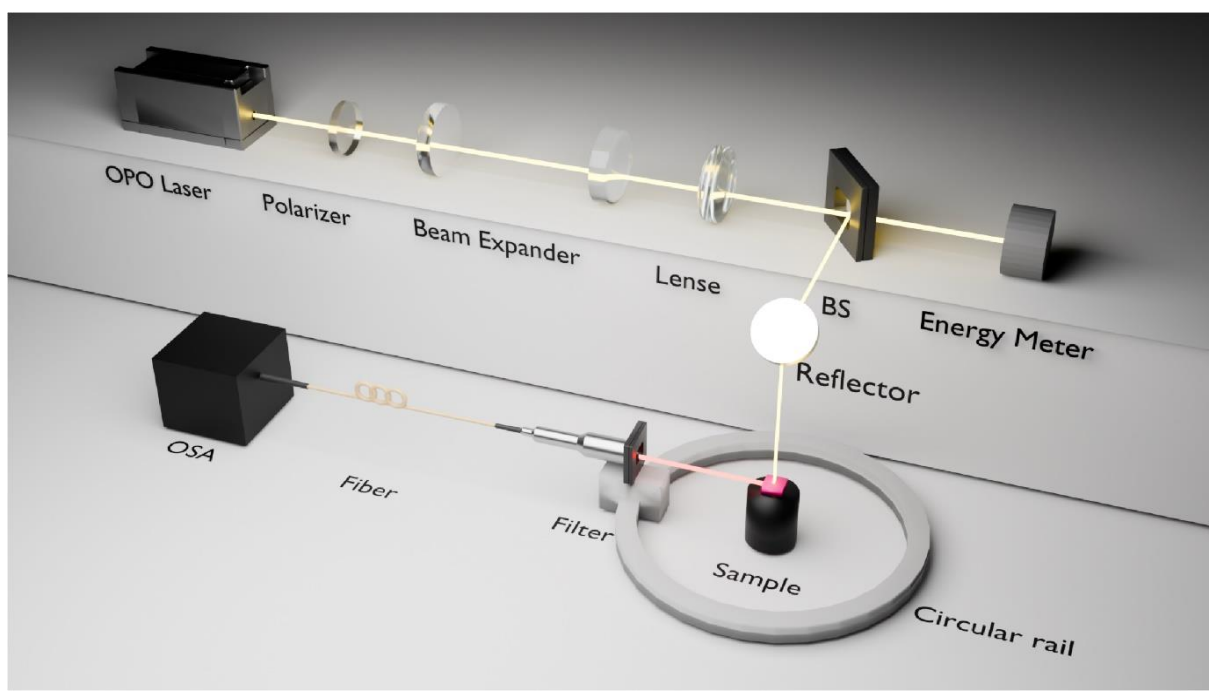

**Fig. S1 Experimental setup for Lasing and directivity measurements.** The Filter and fiber port are fixed on a movable platform that slides freely on a circular rail. OSA: Optical Spectrum Analyzer.

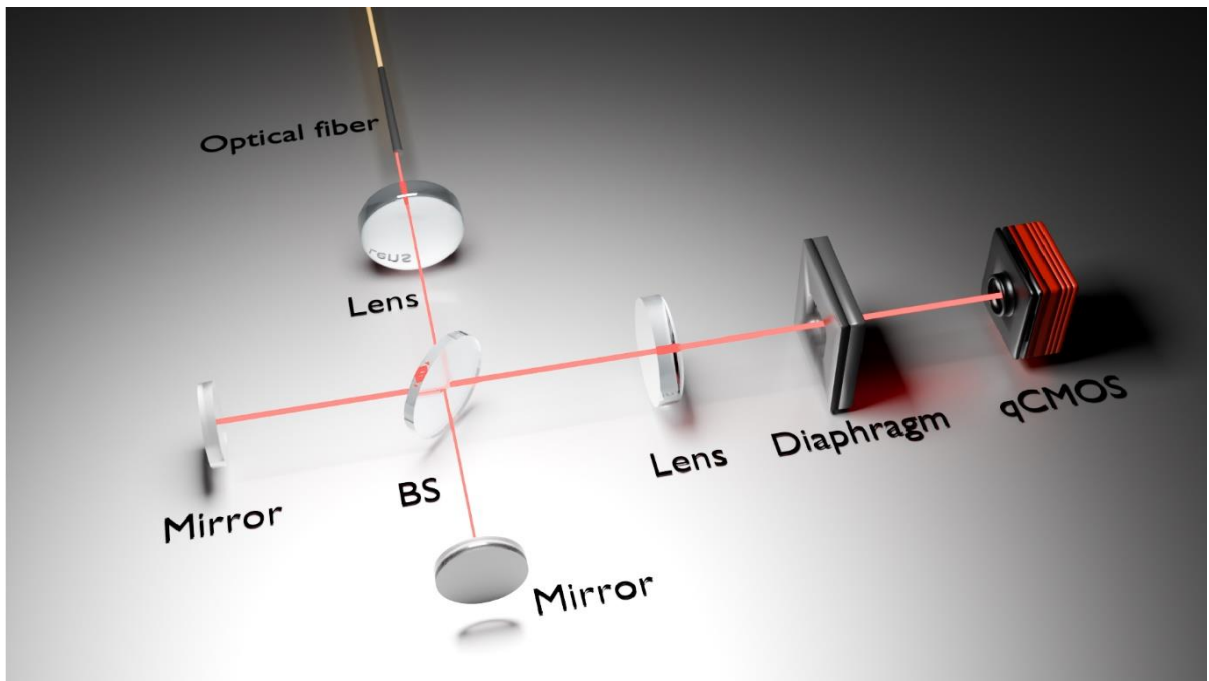

**Fig. S2 Experimental setup for coherence measurements.**

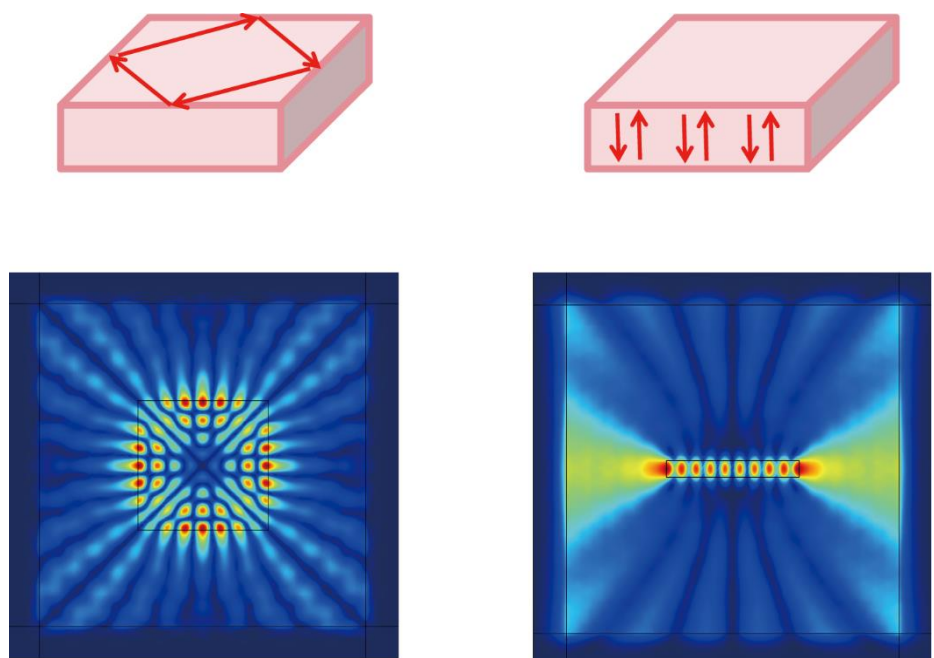

**Fig. S3 Laser mode schematic and diagram**. Left: Whispering gallery mode formed by four sides. Right: Fabry–Pérot mode composed of upper and lower planes.

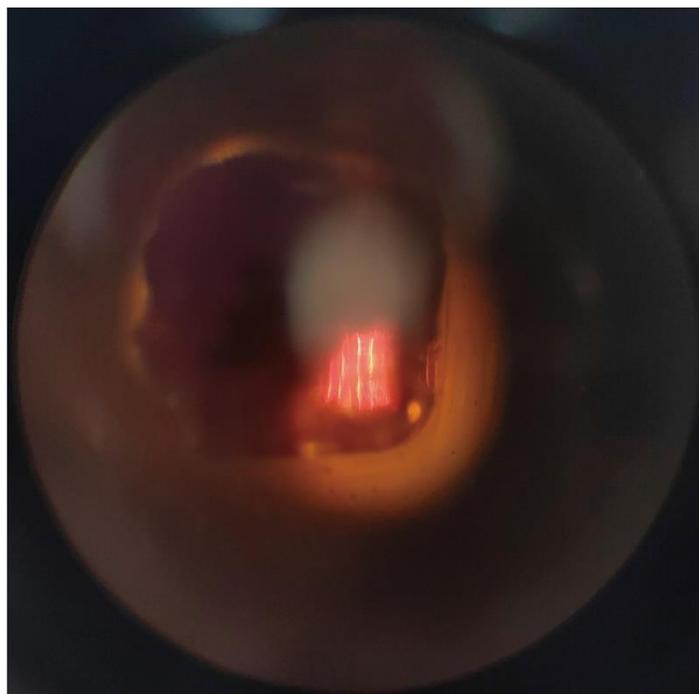

**Fig. S4 An experimental image of finding crystal axes by polarizing microscopy.**

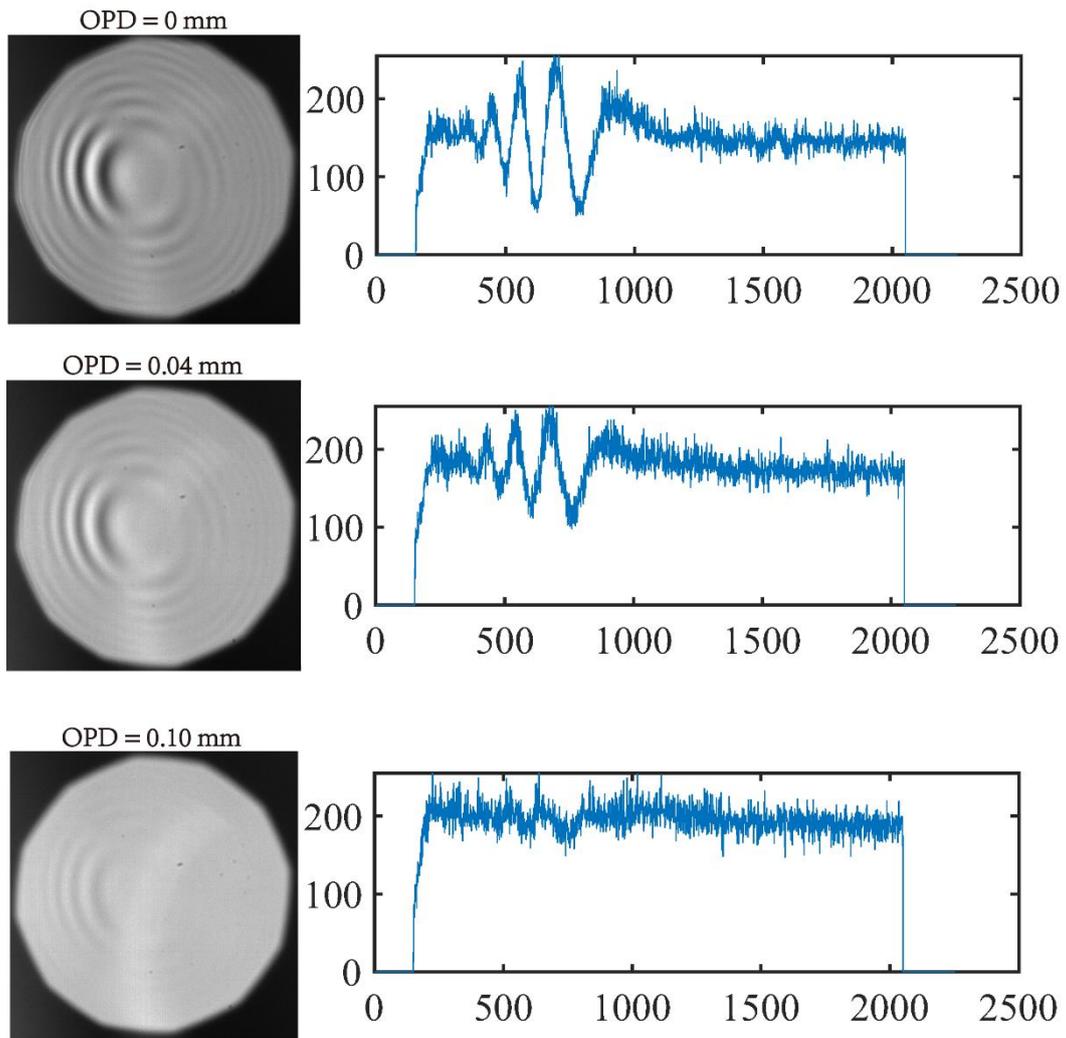

**Fig. S5 Determination of coherence length.** Left: Coherent fringe image (OPD = 0 mm, 0.04 mm, 0.10 mm) taken by the camera. Right: Intensity distribution on the central transverse line of the corresponding image after MATLAB processing.